 \documentclass[final,5p,times,twocolumn,sort&compress]{elsarticle}

\usepackage{amssymb}
\usepackage{lipsum}
\usepackage{graphicx}
\usepackage{dcolumn}
\usepackage{float}
\usepackage{mathptmx, courier, pifont}
\usepackage[scaled=0.92]{helvet}
\usepackage{textcomp}
\usepackage{color}
\usepackage[normalem]{ulem}

\usepackage[dvipsnames]{xcolor}
\usepackage[colorlinks = true,
linkcolor = blue,
urlcolor = blue,
citecolor = blue,
anchorcolor = black]{hyperref} 
\usepackage[all]{hypcap}
\usepackage{makecell}

\usepackage{physics}
\usepackage{numprint}

\journal{Physics Letters B}

\begin{document}

\begin{frontmatter}



\title{Theoretical analysis and predictions for the double electron capture of $^{124}$Xe}


\author[1,2,3]{O. Ni\c{t}escu}
\affiliation[1]{Faculty of Mathematics, Physics and Informatics, Comenius University in Bratislava, 842 48 Bratislava, Slovakia}
\affiliation[2]{International Centre for Advanced Training and Research in Physics, P.O. Box MG12, 077125 Magurele, Romania}
\affiliation[3]{“Horia Hulubei” National Institute of Physics and Nuclear Engineering, 30 Reactorului, POB MG-6, RO-077125 Magurele, Romania}

\author[4,2,3]{S. Ghinescu}
\affiliation[4]{Department of Physics, University of Bucharest, 405 Atomistilor, POB MG-11, RO-077125 Magurele, Romania}

\author[4,2,3]{V.~A.~Sevestrean}

\author[6,2]{M. Horoi}
\affiliation[6]{Department of Physics, Central Michigan University, Mount Pleasant, MI 48859, USA}

\author[1,5]{F. \v{S}imkovic}
\affiliation[5]{Institute of Experimental and Applied Physics, Czech Technical University in Prague,	110 00 Prague, Czech Republic}

\author[2]{S. Stoica}
\ead{sabin.stoica@cifra-c2unesco.ro}

\begin{abstract}

We provide a complete theoretical description of the two-neutrino electron capture in $^{124}$Xe, improving both the nuclear and the atomic structure calculations. We improve the general formalism through the use of the Taylor expansion method, leading to higher order terms in the decay rate of the process. The nuclear part is treated with pn-QRPA and interacting shell model (ISM) methods. The nuclear matrix elements (NMEs) are calculated with the pn-QRPA method with spin restoration by fixing the input parameters so that the experimental decay rate is reproduced, resulting in values significantly lower than in previous calculations. The validity of the pn-QRPA NMEs is tested by showing  their values to be comparable with the ones for double-beta decay with emission of two electrons of $^{128,130}$Te, which have similar pairing features. Within the ISM, we reproduce the total experimental half-life within a factor of two and predict the capture fraction to the KK channel of about 74\%. We also predict the capture fractions to other decay channels and show that for the cumulative decay to the $\rm{KL_{1}}$-$\rm{KO_{1}}$ channels, a capture fraction of about 24\% could be observed experimentally. On the atomic side, calculations are improved by accounting for the Pauli blocking of the decay of innermost nucleon states and by considering all $s$-wave electrons available for capture, expanding beyond the K and L$_1$ orbitals considered in privious studies. We also provide improved atomic relaxation energies of the final atomic states of $^{124}$Te, which may be used as input for background modeling in liquid Xenon experiments. 
\end{abstract}




\end{frontmatter}




\section{Introduction}
\label{sec:Introduction}

The study of double-beta decay (DBD) continues to be of great interest due to its potential to yield novel, essential information about various subjects, including the nuclear structure, the validity of some conservation laws, and still-unknown properties of neutrinos. The experiments have so far largely focused on measuring DBD with the emission of two electrons ($2\nu\beta^-\beta^-$), measured for a dozen nuclei, as a result of their larger $Q$-values compared to other similar transitions and due to their higher natural abundance \cite{GERDAPRL2020,MAJORANA2023,CUPID-0PRL2022,CUOREPRL2022,SNO2022,KamLAND-ZenPRL2023,LEGEND2021}. This heightened attention to DBD is largely attributed to the potential existence of the decay mode without emission of neutrinos ($0\nu\beta^-\beta^-$), whose discovery would open a rich avenue of investigation towards physics beyond the Standard Model (SM) \cite{VergadosRPR2012, DolinskiARNPS2019, AdamsARXIV2022}. 

Besides these transitions, there are three other DBD modes less investigated so far due to their lower transition probabilities, the double-positron emitting ($2\nu/0\nu\beta^+\beta^+$) mode, the single electron-capture with coincident positron emission ($2\nu/0\nu\textrm{EC}\beta^+$) mode and the double electron capture ($2\nu/0\nu\textrm{ECEC}$) mode. However, in the recent past, the ECEC mode has become attractive as well, since in addition to being energetically favored compared to the other two, a resonant transition can occur ($\textrm{R}0\nu\textrm{ECEC}$) if the initial and final states are energetically degenerate, which may increase the transition probability by several orders of magnitude~\cite{BlaumRMP2020}. Therefore, their sensitivity to probe neutrinoless transitions can reach that of the $2\nu\beta^-\beta^-$ decays, so they are considered as an alternative venue to the discovery of neutrinoless decay mode. Double-electron captures with the emission of two neutrinos have been predicted for several isotopes but only measured in $^{78}$Kr, $^{124}$Xe, $^{130}$Ba and $^{132}$Ba~\cite{Belli-P2021}. Particularly, $^{124}$Xe is an interesting case since it has the largest $Q$-value of all the isotopes that can undergo ECEC transitions. It is currently investigated in large experiments together with Dark Matter (DM) and DBD that use detectors with large-volume liquid-xenon \cite{xmass-2018,XENONnT2019,XENON1nTPRC2022,NEXT2021,LUX2020}.

The experimental measurements also require a reliable theoretical support, therefore there are several calculations of the ECEC decay rates and half-lives of $^{124}$Xe~\cite{SuhonenJPG2013,PirinenPRC2015,COELLOPEREZPLB2019}, where the main goal focused on obtaining a more reliable treatment of the nuclear part.  In~\cite{SuhonenJPG2013,PirinenPRC2015}, the half-lives were calculated for the three possible channels, $\beta^+\beta^+$, $\beta^+\rm{EC}$ and ECEC. For the double-capture transitions, only captures of electrons from the K and L$_1$ atomic shells were included in the calculation. The phase space factors (PSFs) were calculated using the formalism from~\cite{DoiPTP1992}, while the nuclear part was calculated with the pn-QRPA method (see \cite{SuhonenJPG2013,PirinenPRC2015} and Refs. therein). An important finding was that the calculated NME values are close for all three channels, so the PSFs are decisive in establishing the magnitude of the half-lives. In~\cite{COELLOPEREZPLB2019}, the $2\nu\rm{ECEC}$ half-lives were calculated using two different methods for the NME calculation, the effective theory (ET) and large-scale Interacting Shell Model (ISM). However, the electron capture processes involve important atomic effects, which were not taken into account in the previous calculations and are relevant for an accurate analysis of the experimental data.

In this work, we provide a complete description of the $2\nu\rm{ECEC}$ in $^{124}$Xe. We improve the general formalism through the use of the Taylor expansion method \cite{Simkovic-2018,Nitescu-U2021}, where we include terms up to the fourth power in the lepton energies in the derivation of the decay rate. Aside from increased rigor, this method allows the definition of new NME ratios, $\xi_{31}^{2\nu\textrm{ECEC}}$ and $\xi_{51}^{2\nu\textrm{ECEC}}$, whose measurements could provide additional insight regarding the interplay between low- and higher-lying states in the intermediate nucleus involved in the NME calculations. Besides this general improvement, we re-examine the calculations of both the atomic and nuclear parts. For the former, we employ the Dirac-Hartree-Fock-Slater (DHFS) self-consistent framework to obtain the electron wave functions. The atomic screening effect is accounted for more rigorously and with better accuracy than in the previous works \cite{DoiPTP1992,KotilaPRC2013,StoicaFP2019}. The diffuse nuclear surface correction, realistic nuclear charge density, and electron exchange-correlation effects are also taken into account within our atomic structure calculations. We further improve the PSF calculation by accounting for the Pauli blocking of the decay of innermost nucleon states and by considering all $s$-wave electrons available for capture, expanding beyond the K and L$_1$ orbitals considered in prior studies. We also compute the atomic de-excitation energies within the DHFS framework, which, along with the capture fractions, can be used as input for background modeling in liquid xenon experiments.

For the nuclear part we use the ISM~\cite{HoroiStoicaBrown2007,SenkovHoroiBrown2014,SenkovHoroi2016,NeacsuHoroi2015,NeacsuHoroi2016,physics4040074} and the pn-QRPA with isospin restoration~\cite{Sim13} methods. The 2$\nu$ECEC NMEs, as the 2$\nu\beta\beta$ NMEs, can only be calculated reliably if one includes the sum on the excited $1^+$ states of the intermediate nucleus (see Eqs. (\ref{NME})), which is possible in the ISM and pn-QRPA approaches. 
Methods that cannot calculate the full sum on the intermediate $1^+$ states rely on closure approximation or single-state dominance approximation, which rarely provides reliable results. In the ISM one can replace the direct sum on intermediate states with a strength function approach, which converges much quicker to the exact result. A full description of the strength function approach within the ISM framework can be found in section 4 of Ref. \cite{physics4040074}. This method is extended in this work to higher powers in the denominators (see Eqs. (\ref{NME})) necessary for the Taylor expansion terms. The new ISM NMEs and the improved PSFs are used to predict capture fractions for 2$\nu$ECEC channels that were not yet observed. In the analysis we considered two often-used effective Hamiltonians, which provide consistent nuclear structure results, thus reassuring of the reliability of the calculated NMEs and of the newly predicted capture fractions.

We also compute the NMEs with the pn-QRPA method with isospin restoration by fixing the particle-particle strength parameter to reproduce the experimental half-life. We found that our NME value is much smaller when compared with the results of previous pn-QRPA calculations \cite{SuhonenJPG2013,PirinenPRC2015}.

We organize the manuscript \textit{}as follows: Section~\ref{sec:Formalism} provides a description of the improved $2\nu$ECEC capture formalism, including the most relevant equations. Section~\ref{sec:ResultsAndDiscussions} presents our results of the $2\nu$ECEC process of $^{124}$Xe, including the refined PSFs, the atomic relaxation energies, the NMEs obtained with the ISM and pn-QRPA, and the total and partial half-lives predictions. The conclusions are summarized in Section~\ref{sec:Conclusions}.


\section{Improved two-neutrino double electron capture formalism}
\label{sec:Formalism}
We improve the $2\nu\textrm{ECEC}$ formalism through the use of a Taylor expansion method. Following the same steps as for $2\nu\beta^-\beta^-$ decay \cite{Simkovic-2018,Nitescu-U2021} and including terms up to the fourth power in the lepton energies, we obtain the total inverse half-life of the $2\nu\textrm{ECEC}$ process in a similar form:
\begin{align}\label{ttt}
  \begin{aligned}
  \left[T^{2\nu\rm{ECEC}}_{1/2}\right]^{-1}&=\left(g^{\rm eff}_A\right)^4\left|M^{2\nu\rm{ECEC}}_{GT-1}\right|^2\left\{ G_0^{2\nu\rm{ECEC}}\right.\\
  &+\left.\xi^{2\nu\rm{ECEC}}_{31}G_2^{2\nu\rm{ECEC}} +\frac{1}{3}\left(\xi^{2\nu\rm{ECEC}}_{31}\right)^2G_{22}^{2\nu\rm{ECEC}}\right.\\
  &+\left.\left[\frac{1}{3}\left(\xi^{2\nu\rm{ECEC}}_{31}\right)^2+\xi^{2\nu\rm{ECEC}}_{51}\right]G_{4}^{2\nu\rm{ECEC}}\right\},
  \end{aligned}
\end{align}
 The partial inverse half-life for $2\nu\textrm{xy}$ process, where the atomic electrons are only captured from shells $\rm{x}$ and $\rm{y}$, can be written as:
\begin{align}\label{tkk}
  \begin{aligned}
  \left[T^{2\nu\rm{xy}}_{1/2}\right]^{-1}=&\left(2-\delta_{\rm{xy}}\right)\left(g^{\rm eff}_A\right)^4\left|M^{2\nu\rm{ECEC}}_{GT-1}\right|^2\left\{ G_0^{2\nu\rm{xy}}+\right.\\
  &+\left.\xi^{2\nu\rm{ECEC}}_{31}G_2^{2\nu\rm{xy}} +\frac{1}{3}\left(\xi^{2\nu\rm{ECEC}}_{31}\right)^2G_{22}^{2\nu\rm{xy}}+\right.\\
  &+\left.\left[\frac{1}{3}\left(\xi^{2\nu\rm{ECEC}}_{31}\right)^2+\xi^{2\nu\rm{ECEC}}_{51}\right]G_{4}^{2\nu\rm{xy}}\right\}.
  \end{aligned}
\end{align} 
Here, $g^{\rm eff}_A$ is an effective axial coupling constant and the parameters:
\begin{equation}\label{xi1-3}
  \xi^{2\nu\rm{ECEC}}_{31}=\frac{M^{2\nu\rm{ECEC}}_{GT-3}}{M^{2\nu\rm{ECEC}}_{GT-1}}, \hspace{0.5cm} \xi^{2\nu\rm{ECEC}}_{51}=\frac{M^{2\nu\rm{ECEC}}_{GT-5}}{M^{2\nu\rm{ECEC}}_{GT-1}}
\end{equation}
are NME ratios arising from the Taylor expansion, with
\begin{align} \label{NME}
  \begin{aligned}
    M^{2\nu\rm{ECEC}}_{GT-1} &= \sum_n M^{2\nu}_{GT}(n)\frac{m_e}{E_n(1^+)-(E_i+E_f)/2}, \\
    M^{2\nu\rm{ECEC}}_{GT-3} &= \sum_n M^{2\nu}_{GT}(n) \frac{4~ m_e^3}{\left(E_n(1^+) - (E_i+E_f)/2\right)^3},\\
    M^{2\nu\rm{ECEC}}_{GT-5} &= \sum_n M^{2\nu}_{GT}(n) \frac{16~ m_e^5}{\left(E_n(1^+) - (E_i+E_f)/2\right)^5}.
  \end{aligned}
\end{align}
The summations are over all $1^+$ states of the intermediate nucleus and $M^{2\nu}_{GT}(n)$ matrix elements depend on the $n^{\rm th}$ $1^+$ intermediate state, with energy $E_n(1^+)$, as well as on the ground states $|0^+_i\rangle$ and $|0^+_f\rangle$ of the initial and final nuclei, with the energies $E_i$ and $E_f$:
\begin{equation}
	\label{gttran} 
	M^{2\nu}_{GT}(n)=\langle 0^+_f\|\sum_{m}\tau^-_m\sigma_m\| 1^+_n\rangle \langle 1^+_n\|\sum_{m}\tau^-_m\sigma_m\| 0^+_i\rangle,
\end{equation}
where $\tau^-_m$ is the isospin-lowering operator transforming a proton into a neutron, and $\sigma_m$ is the nucleon spin operator.

The PSFs entering the inverse half-lives are given by
\begin{align}
\begin{aligned}
\label{eq:PSFTaylor}
G_N^{2\nu\rm{ECEC}}&=\frac{m_e(G_{F}\left|V_{ud}\right|m_e^2)^4}{2\pi^3\ln{(2)}}\frac{1}{m_e^{5}}\sum_{\rm{x,y}}\mathcal{B}_\mathrm{x}^2\mathcal{B}_\mathrm{y}^2\mathcal{I}_{N,\rm{xy}}\\
          &=\sum_{\rm{x,y}}G_N^{2\nu\rm{xy}},
\end{aligned}
\end{align}
where $G_F$ is the Fermi coupling constant, $V_{ud}$ the first element of the Cabibbo-Kobayashi-Maskawa (CKM) matrix and the summations run over all occupied atomic shells of the initial atom. The functions $\mathcal{I}_{N,\rm{xy}}$ depend on the total energies of the electrons from the initial atom's orbitals $\rm{x}$ and $\rm{y}$, denoted as $e_\textrm{x}$ and $e_\textrm{y}$, respectively. Their expressions can be obtained from Eqs. A4 and A5 of Ref. \cite{Nitescu-U2021} by making the replacements, $E_{e_1}\rightarrow-e_\textrm{x}$ and $E_{e_2}\rightarrow-e_\textrm{y}$. The localization probability of an electron from shell $\rm{x}$ inside the nucleus, is given by:
\begin{equation}
    \label{eq:LocalizationProb}
	\mathcal{B}^2_{\rm{x}}=\frac{1}{4\pi m_e^3}\left[\langle g_{\rm{x}}\rangle^2+\langle f_{\rm{x}}\rangle^2\right],
\end{equation}
where
\begin{equation}
    \langle g_{\rm{x}}\rangle=\frac{\int g_{\rm{x}}(r)\rho(r)r^2dr}{\int \rho(r)r^2dr}\hspace{1cm} \langle f_{\rm{x}}\rangle=\frac{\int f_{\rm{x}}(r)\rho(r)r^2dr}{\int \rho(r)r^2dr}
\end{equation}
Here, $g_{\rm{x}}(r)$ and $f_{\rm{x}}(r)$ are, respectively, the large- and small-component of the radial wave functions for the bound electron in shell x, and the nuclear charge distribution normalized to $Z$ is given by:
\begin{equation}
    \rho(r)=\frac{1}{1+e^{\left(r-c_{\rm rms}\right)/a}}
\end{equation}
where the surface thickness $a = 0.545$ fm was used and we found that $c_{\rm{rms}}= 5.569$ fm reproduces the experimental mean square radius, $\sqrt{\langle r^2 \rangle}=4.7661$ fm, for $^{124}$Xe \cite{Angeli-ADNDT2013}. The average of the bound wave function, weighted with the nuclear charge distribution, accounts for the Pauli blocking of the decay of innermost nucleon states while preserving the intuitive picture of the electron being captured on the nuclear surface.

\section{Results and Discussions for the $2\nu\rm{ECEC}$ decay of $^{124}\rm{Xe}$}
\label{sec:ResultsAndDiscussions}

\subsection{Atomic structure description, PSF calculation and atomic relaxation energies}
\label{subsec:ElectronWaveFunctions}

Accurate computation of the PSFs for the $2\nu\textrm{ECEC}$ process in $^{124}$Xe and the determination of atomic relaxation energies for the final atom, $^{124}$Te, requires precise atomic structure calculations for these atomic systems. For this purpose, we employed the Dirac-Hartree-Fock-Slater (DHFS) self-consistent framework. The nuclear, electronic and exchange components of the DHFS potential and the convergence of the method were described in detail in \cite{NitescuPRC2023,SevestreanPRA2023}, based on the \textsc{RADIAL} subroutine package \cite{Salvat-CPC2019}, that we also employ in this paper. In a relativistic framework, the shells referred to as $\rm{x}$ or $\rm{y}$, in previous sections, can be uniquely identified by the state $n\kappa$, where $n$ represents the principal quantum number and $\kappa$ corresponds to the relativistic quantum number.

Usually, in the $2\nu$ECEC process, it is considered that the electrons are captured from K shell and L$_1$ subshell \cite{DoiPTP1992,KotilaPRC2013,StoicaFP2019}. In particular for $^{124}$Xe double electron capture process, even the electrons from $O$ shell have a small but non-zero probability to be captured. In this paper, we assume that the neutrinos are emitted with anti-parallel spins and that the electrons can be captured from all occupied shells of the initial atom, so the summations in Eq.~\ref{eq:PSFTaylor} extend up to $O_1$ subshell. We have considered only the subshells with $\kappa=-1$. Captures in which not all four leptons are in $s$-wave states are strongly suppressed by additional terms in the NMEs \cite{DoiPTP1992}. This may influence the interpretation of experimental results. For example, in \cite{XENON1nTPRC2022}, the signal model uses captures from all combinations of subshells. In our model, each subshell has only two electrons, hence there are two pairs of electrons that can be captured for any given $\rm{x} \neq \rm{y}$ subshells. This is accounted for in Eq.~(\ref{eq:PSFTaylor}) by not requiring $\rm{x}<\rm{y}$.

The results of the PSFs, obtained using the Taylor expansion formalism for the $2\nu\text{ECEC}$ process of $^{124}$Xe, are summarized in Table~\ref{tab:PSFTaylor} according to Eq.~(\ref{eq:PSFTaylor}). The first line corresponds to the total PSFs, while the subsequent lines display results for the first few most probable channels. As anticipated, with the order of the Taylor expansion increasing, the values of the PSFs decrease. This trend aligns with data for the case of $2\nu\beta^-\beta^-$ decay \cite{Simkovic-2018,Nitescu-U2021}. At first glance, one can compare the first entry in Table~\ref{tab:PSFTaylor} with the older PSF, e.g., $17200\times10^{-24}$ yr$^{-1}$ from Ref.~\cite{KotilaPRC2013}, but the comparison must be approached with care. Firstly, the result from Ref.\cite{KotilaPRC2013} must be combined with the closure NME, while our values are designed for the Taylor expansion NMEs. Secondly, the older value includes only contributions from K and L$_1$ captures, whereas we take into account all possible captures of $^{124}$Xe. As a general result \cite{Nitescu-U2024}, we found that the DHFS screening decreases the total PSF when compared with the Thomas-Fermi screening from \cite{KotilaPRC2013,StoicaFP2019}. In contrast, Pauli blocking effects and the inclusion of captures beyond K and L$_1$ increase its value.

\begin{table}[h]
\centering
\begin{tabular}{c|r|r|r|r}
      & $N=0$ & $N=2$ & $N=22$ & $N=4$ \cr 
		\hline
    $G_N^{2\nu\rm{ECEC}}$ & $18332.0  $ & $9802.2  $  & $3056.1  $ & $6116.3  $\\
    $G_N^{2\nu\rm{KK}}$  & $13605.3  $ & $7230.0  $ & $2241.2  $ & $4482.4  $\\
    $G_N^{2\nu\rm{KL_1}}$  & $1741.9  $ & $945.8  $ & $298.9  $ & $599.4  $\\
    $G_N^{2\nu\rm{KM_1}}$  & $355.4  $ & $193.6  $ & $61.3  $ & $123.1  $\\
    $G_N^{2\nu\rm{KN_1}}$  & $78.6  $ & $42.8  $ & $13.6  $ & $27.3  $\\
    $G_N^{2\nu\rm{KO_1}}$  & $11.8  $ & $6.4  $ & $2.0  $ & $4.1  $\\
    $G_N^{2\nu\rm{L_1L_1}}$  & $222.9  $ & $123.4  $ & $39.9  $ & $79.8  $\\
    $G_N^{2\nu\rm{L_1M_1}}$  & $45.5  $ & $25.3  $ & $8.2  $ & $16.4  $\\
\end{tabular}
\caption{\label{tab:PSFTaylor} The evaluated PSFs (Eq.~\ref{eq:PSFTaylor}) units of $10^{-24}\textrm{yr}^{-1}$ for the total $2\nu\textrm{ECEC}$ process (first line) and the partial $2\nu\textrm{xy}$ processes (the following lines) of $^{124}$Xe. We have used $Q=2856.73$ keV \cite{QValue124Xe-PRC2012}.}
\end{table}

After the capture takes place, the final atom is left in an excited state. Experimentally, the detection of the de-excitation energies (in the form of X-rays and Auger electrons) is used to identify 2$\nu$ECEC events, but it can also interfere with other events from experiments where the decay of $^{124}$Xe is an unavoidable source of background. The atomic relaxation energy subsequent to the $2\nu \textrm{xy}$ process can be written in terms of the total electron binding energies of $^{124}$Te in the ground state and in the excited state with holes in x and y orbitals as
\begin{equation}
	\label{eq:AtomicRelaxationEnergy}
	R_{\rm xy}=B_{\rm g.s.}(^{124}\textrm{Te})-B_{\rm xy}(^{124}\textrm{Te}).
\end{equation}
The energies obtained from the DHFS framework are displayed in Table~\ref{tab:RelaxationEnergies}. According to \cite{SevestreanPRA2023}, our $R_{\mathrm{xy}}$ estimates are expected to have an accuracy better than 1\%. Considering the current experimental resolution of a few keV \cite{XENON1nTPRC2022}, our estimation is a viable option for establishing the background peak position in liquid xenon experiments \cite{XENON1nTPRC2022,DARWIN-JCAP2016,DARWIN-JPG2022}.

\subsection{The ISM evaluation of the NMEs}
\label{ssec:NMEEvaluationSM}

The nuclear matrix elements (NMEs), Eq. (\ref{NME}), have been calculated similarly as in Ref. \cite{COELLOPEREZPLB2019} but with few significant differences: (i) no single state dominance (SSD) was assumed \cite{physics4040074}, (ii) two effective Hamiltonians were employed, including GCN5082 \cite{COELLOPEREZPLB2019} (name shortened to GCN in the tables below), but also SVD \cite{Chong2012}, (iii) unique quenching factors ($q_H$) for the GT $\tau^- \sigma$ operator, were employed for each effective Hamiltonian, which describe well the two-neutrino double beta decay data of $^{136}$Xe for each case, namely $q_{GCN}=0.4$ \cite{COELLOPEREZPLB2019} and $q_{SVD}=0.7$ \cite{NeacsuHoroi2015,NeacsuHoroi2016}, and (iv) up to four nucleons were excited from the lower $g_{7/2}d_{5/2}$ orbitals into the higher $jj55$-space orbitals.

As in Ref. \cite{COELLOPEREZPLB2019} we calculate the NME in the $jj55$ model space consisting of the $0g_{7/2}$, $1d_{5/2}$, $1d_{3/2}$, $2s_{1/2}$, and the $0h_{11/2}$ orbitals for both protons and neutrons. The shell model calculations in the full model space for all nuclei involved in the transition are hindered by the high dimensional basis needed, especially for $^{124}$Xe. As in Ref. \cite{COELLOPEREZPLB2019}, we rely on truncations, by promoting nucleons from the lower $g_{7/2}d_{5/2}$ orbitals into the higher $d_{3/2}s_{1/2}h_{11/2}$ orbitals. In this work we go beyond the truncations used in Ref. \cite{COELLOPEREZPLB2019}, by allowing up to four nucleon excitations. In addition, we go beyond the one state dominance approximation used in Ref. \cite{COELLOPEREZPLB2019} by performing a full summation on the intermediate $1^+$ states in Eq. (\ref{NME}) \cite{HoroiStoicaBrown2007} using a strength function approach described in detail in section 4 of Ref. \cite{physics4040074}. The full summation reduces the NME for the GCN5082 Hamiltonian by about 25\%. In addition, as mentioned in the previous paragraph, we don't assume a range of unjustified quenching factors, but we use the same quenching factors that describe reasonably well the NME for the two-neutrino double beta decay of $^{136}$Xe calculated in the same model space and using the same effective Hamiltonians.

\begin{figure}[h]
		\centering
		\includegraphics[width=0.48\textwidth]{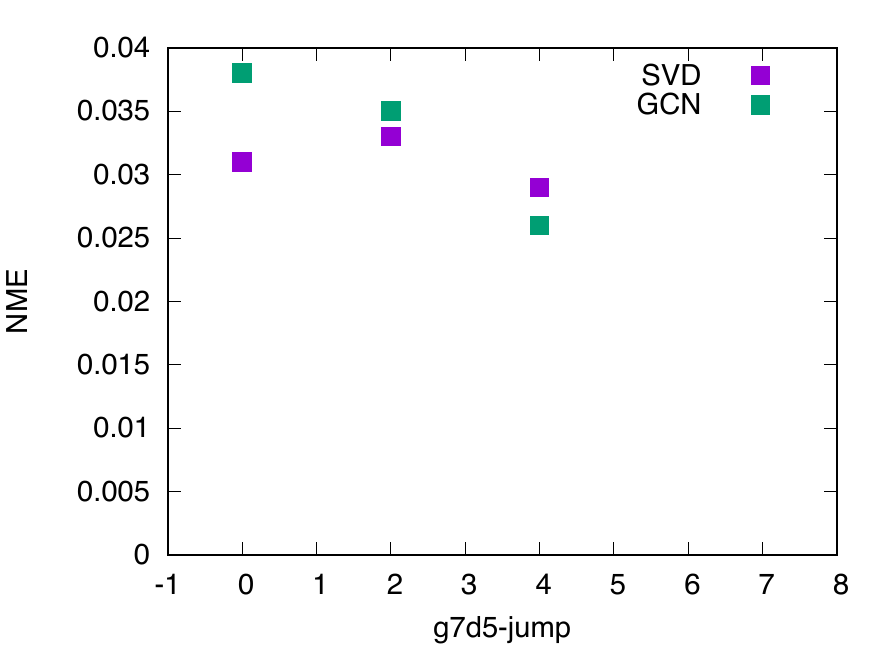}
		\caption{The $M^{2\nu ECEC}_{GT-1}$ NME as a function of the number of nucleons that were allowed to get excited from the lower $g_{7/2}d_{5/2}$ orbitals.}
 \label{nme-ecec}
\end{figure}

The results for the dominant matrix elements, $M_{GT-1}^{2\nu \rm{ECEC}}$, are presented in Fig. \ref{nme-ecec}, where g7d5-jump represents the number of nucleons that were allowed to get excited (jump) from the lower $g_{7/2}d_{5/2}$ orbitals into the higher $jj55$-space orbitals, maximum being 14. One can see that both effective Hamiltonians provide similar results, which, although not yet converged, do not significantly vary, especially for the SVD effective Hamiltonian. We also calculated the other two nuclear matrix elements in Eq. (\ref{NME}), $M_{GT-3}^{2\nu \rm{ECEC}}$ and $M_{GT-5}^{2\nu \rm{ECEC}}$, using similar techniques by extending the approach presented in section 4 of Ref. \cite{physics4040074}. In these calculations the Gamow-Teller $\tau^- \sigma$ operator in Eq. (\ref{gttran}) was quenched by the $q_H$ factors given in the previous paragraph ($q_{SVD}=0.7, q_{GCN}=0.4$). The ISM point of view is that this quenching effect is due to the renormalization of the $\tau^- \sigma$ operator in reduced model spaces \cite{PhysRevC.100.014316,particles4040038,Gysbers2019-tm},  while the axial coupling constant remains that for free nucleons. Therefore in ISM the "renormalized" $\tau^- \sigma$ operator in Eq. (\ref{gttran})  is  $ q_H \tau^- \sigma$, and $g^{\textrm{eff}}_A$ in Eqs. (\ref{ttt})-(\ref{tkk}) is  $g_A=1.276$~\cite{PhysRevLett.122.242501}. The NME results are summarized in Table \ref{nmes}, where "jump" is short for "g7d5-jump". These results are further used to extract the $\xi_{31}^{2\nu\textrm{ECEC}}$ and $\xi_{51}^{2\nu\textrm{ECEC}}$ parameters, presented in Table \ref{xis}, which are needed for calculating the Taylor expansion corrections to the decay half-lives, Eqs. (\ref{ttt}-\ref{tkk}). One should emphasize the $M_{GT-3}^{2\nu \rm{ECEC}}$ and $M_{GT-5}^{2\nu \rm{ECEC}}$ NMEs exhibit similar behavior as that presented in Fig. \ref{nme-ecec} by the $M_{GT-1}^{2\nu \rm{ECEC}}$ NMEs.

\begin{table}[h]
\centering
\begin{tabular}{c|c|c|c|c}
Model & NME type & jump=0 & jump=2 & jump=4\cr 
		\hline
 \hline 
SVD  &  $M_{GT-1}^{2\nu \rm{ECEC}}$ & 0.0305 & 0.0329 & 0.0291\\
        &  $M_{GT-3}^{2\nu \rm{ECEC}}$ & 0.0066 & 0.0087 & 0.0064\\
  &  $M_{GT-5}^{2\nu \rm{ECEC}}$ & 0.0018 & 0.0026 & 0.0018\\
  \hline
GCN &  $M_{GT-1}^{2\nu \rm{ECEC}}$ & 0.0379 & 0.0354 & 0.0264\\
  &  $M_{GT-3}^{2\nu \rm{ECEC}}$ & 0.0108 & 0.0092 & 0.0057\\
  &  $M_{GT-5}^{2\nu \rm{ECEC}}$ & 0.0032 & 0.0027 & 0.0016\\
\end{tabular} 
\caption{\label{nmes} ISM results for the NMEs of Eq. (\ref{NME}) (see text for details).}
\end{table}

\begin{table}[h] 
\centering
\begin{tabular}{c|c|c|c|c}
Model & $\xi_i^{2\nu\textrm{ECEC}}$ & jump=0 & jump=2 & jump=4\cr 
		\hline
 \hline 
SVD &  $\xi _{31}^{2\nu\textrm{ECEC}}$ & 0.216 & 0.264 & 0.220\\
	&  $\xi _{51}^{2\nu\textrm{ECEC}}$ & 0.059 & 0.079 & 0.062\\
  \hline
GCN &  $\xi _{31}^{2\nu\textrm{ECEC}}$ & 0.285 & 0.260 & 0.216\\
  &  $\xi _{51}^{2\nu\textrm{ECEC}}$ & 0.084 & 0.076 & 0.061\\
  
\end{tabular}
\caption{\label{xis} ISM results for the $\xi _i$ parameters of Eq. (\ref{xi1-3}) (see text for details).}
\end{table}

\subsection{The NME evaluation in pn-QRPA}
\label{ssec:NMEEvaluationQRPA}

The nuclear matrix elements $M^{2\nu \rm{ECEC}}_{GT-1}$, $M^{2\nu \rm{ECEC}}_{GT-3}$ and $M^{2\nu \rm{ECEC}}_{GT-5}$ are calculated using the pn-QRPA with isospin restoration \cite{Sim13}. They were obtained considering the same large model space (23 subshells of N = 0–5 oscillator shells with the addition of $i_{11/2}$ and $i_{13/2}$) and mean fields as in Ref. \cite{Sim13} for double-beta of decay $^{128,130}$Te and  $^{136}$Xe. Pairing and residual interactions are derived from the same modern realistic nucleon-nucleon potentials, namely from charge-dependent Bonn potential (CD-Bonn) and the Argonne V18 potential. When solving the BCS pairing equations, the strengths of pairing interactions are slightly renormalized so that experimental pairing gaps are correctly reproduced \cite{Sim13}. The pn-QRPA equations contain three renormalization adjustable parameters: $g_{ph}$ for particle-hole interaction, $g_{pp}^{T=1}$ and $g_{pp}^{T=0}$ for isovector and isoscalar parts of the particle-particle interaction. While $g_{ph} = 1.0$ is typically used \cite{Sim13}, $g_{pp}^{T=1}$ is fixed by the requirement that the $2\nu\beta^-\beta^-$ Fermi matrix element vanishes, as it should. $g^{T=0}_{pp}$ is adjusted so that the half-life of the $2\nu$ECEC of $^{124}$Xe is correctly reproduced for each considered value of $g^{\rm eff}_A$. Unlike earlier QRPA calculations of $2\nu\beta^-\beta^-$ NMEs, the BCS overlap of the initial and final vacua is taken into account, leading to a reduction by factor 0.828 \cite{Sim04}.

\begin{figure}[h]
\centering
\includegraphics[width=0.48\textwidth]{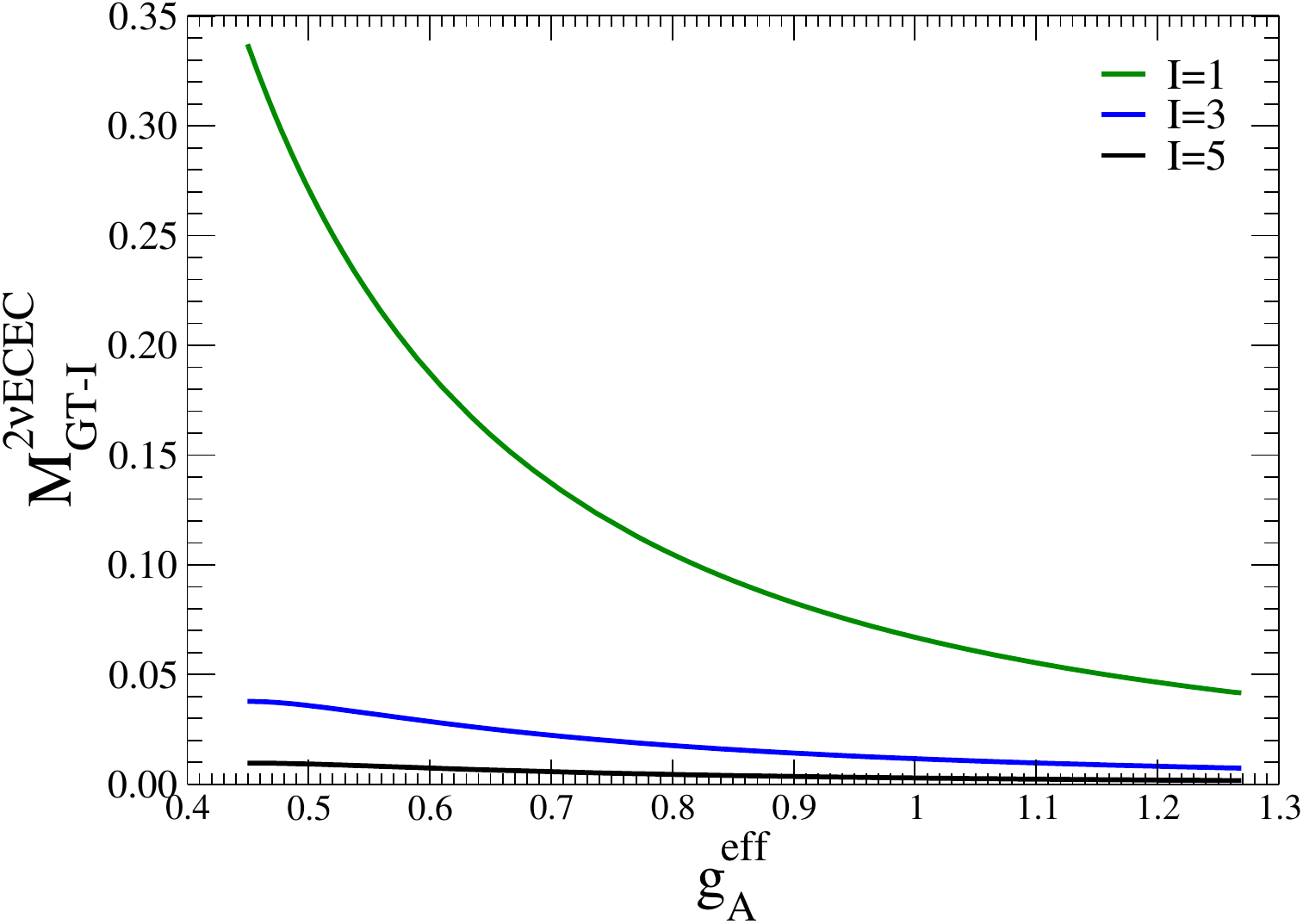}
\caption{Dependence of the matrix elements $M^{2\nu \rm ECEC}_{GT-1}$, $M^{2\nu \rm ECEC}_{GT-3}$, and $M^{2\nu \rm{ECEC}}_{GT-5}$
  on the effective value of axial-vector coupling constant $g_A^{\rm eff}$ for the case the calculated $2\nu$ECEC half-life
is the same as its experimental value, i.e., $T^{2\nu \textrm{ECEC}}_{1/2}(^{124}\textrm{Xe}) = 1.1\times 10^{22}$ yrs \cite{XENON1nTPRC2022}. Results were obtained within the QRPA with the restoration of the isospin by assuming the realistic Argonne V18 nucleon-nucleon potential.}\label{qrpaNMEs}
\end{figure}

\begin{table*}[!t]
\caption{
  Nuclear matrix elements
  $M^{M}_{\rm GT-K}$ (M = $2\nu$ECEC and $2\nu\beta^-\beta^-$, K=1,3 and 5) for $2\nu$ECEC of $^{124}$Xe and $2\nu\beta^-\beta^-$-decay of $^{128,130}$Te calculated within pn-QRPA  with isospin restoration. $\Delta_p$ and $\Delta_n$ denote experimental proton and neutron pairing gap, respectively. $g^{\rm eff}_A$ is the effective axial-vector coupling constant. $A$ and $B$ stand for the Argonne V18 and CD-Bonn nucleon-nucleon potential, respectively. The fixed value of the particle-particle strength parameter in the isovector channel $g^{T=1}_{pp}$ is 0.9626 (0.8884), 0.9672 (0.8934) and 0.9893 (0.9151) for $2\nu$ECEC of $^{124}$Xe, $2\nu\beta^-\beta^-$-decay of $^{128}$Te and $^{130}$Te, respectively, by exploiting the realistic  Argonne V18 (CD-Bonn) nucleon-nucleon potential. $g^{T=0}_{pp}$ is the renormalization constant of particle-particle neutron-proton interaction in the isoscalar channel, which is fitted from the requirement that the calculated $2\nu$ECEC or $2\nu\beta^-\beta^-$ half-life is the same as its experimental value. $T^{2\nu \textrm{ECEC}}_{1/2}(^{124}\textrm{Xe}) = 1.1\times 10^{22}$ yrs \cite{XENON1nTPRC2022}, $T^{2\nu\beta^-\beta^-}_{1/2}(^{128}\textrm{Te}) = 2.3\times 10^{24}$ yrs \cite{Barabash19} and $T^{2\nu\beta^-\beta^-}_{1/2}(^{130}\textrm{Te}) = 7.71\times 10^{20}$ yrs \cite{Cuore23} are considered.
\label{tab:qrpa}}
\renewcommand{\arraystretch}{1.2}
\begin{tabular}{cccccccccccccccc}\hline\hline
  Nucl. tran. & & $\Delta^{\rm Te}_p $ & $\Delta^{\rm Te}_n$ & $\Delta^{\rm Xe}_p$ & $\Delta^{\rm Xe}_n$ & & $g^{\rm eff}_A$ & pot. &  $g^{T=0}_{pp}$ & $M_{\rm GT-1}$ &   $M_{\rm GT-3}$ &   $M_{\rm GT-5}$ & & $\xi_{31}$ & $\xi_{51}$ \\\hline
 $^{124}$Xe$\rightarrow ^{124}$Te & & 1.249 & 1.344 & 1.357 & 1.339 & & 0.80  & A & 0.706 & 0.1044 & 0.018 & 0.0045 & & 0.169 & 0.043  \\  
                                 & &       &       &       &       & &      & B & 0.635 & 0.1050 & 0.019 & 0.0049 & & 0.181 & 0.047  \\
                                 & &       &       &       &       & & 1.00 & A & 0.737 & 0.0662 & 0.016 & 0.0029 & & 0.175 & 0.043  \\
                                 & &       &       &       &       & &      & B & 0.664 & 0.0677 & 0.014 & 0.0035 & & 0.201 & 0.051  \\
                                 & &       &       &       &       & & 1.27 & A & 0.7542 & 0.0416 & 0.0074 & 0.0017 & & 0.178 & 0.041  \\
                                 & &       &       &       &       & &      & B & 0.6814 & 0.0409 & 0.0094 & 0.0023 & & 0.231 & 0.057 \\
 $^{128}$Te$\rightarrow ^{128}$Xe & & 1.129 & 1.280 & 1.319 & 1.265 & & 0.80  & A & 0.7432 & 0.0631 & 0.012  & 0.0038 & & 0.196 & 0.060  \\  
                                 & &       &       &       &       & &      & B & 0.6676 & 0.0631 &  0.014 & 0.0043 & & 0.219 & 0.068 \\
                                 & &       &       &       &       & & 1.00 & A & 0.7645 & 0.0403 &  0.0081 & 0.0023 & & 0.200 & 0.057  \\
                                 & &       &       &       &       & &      & B & 0.6886 & 0.0404 & 0.0098 & 0.0030 & & 0.244 &  0.073  \\
                                 & &       &       &       &       & & 1.27 & A & 0.7773 & 0.0251 & 0.0051 & 0.0013 & & 0.202 &  0.051  \\
                                 & &       &       &       &       & &      & B & 0.7013 & 0.0250 & 0.0070 & 0.0020 & & 0.282 & 0.081 \\
 $^{130}$Te$\rightarrow ^{130}$Xe & & 1.058 & 1.179 & 1.297 & 1.248 & & 0.80  & A & 0.7490 & 0.0444 & 0.0082 & 0.0023 & & 0.185 & 0.052   \\  
                                 & &       &       &       &       & &      & B & 0.6726 & 0.0443  & 0.0093 & 0.0027 & & 0.209 & 0.060   \\
                                 & &       &       &       &       & & 1.00 & A & 0.7680 & 0.0284 & 0.0054 & 0.0014 & & 0.192 & 0.049  \\
                                 & &       &       &       &       & &      & B & 0.6915 & 0.0282 & 0.0067 & 0.0018 & & 0.237 & 0.065  \\
                                 & &       &       &       &       & & 1.27 & A & 0.7797 & 0.0177 & 0.0035 & 0.0008 & & 0.199 & 0.044   \\
                                 & &       &       &       &       & &      & B & 0.7031 & 0.0174 & 0.0049 & 0.0013 & & 0.281 & 0.072   \\\hline\hline
\end{tabular}
\end{table*}

In Fig. \ref{qrpaNMEs}, the calculated $2\nu$ECEC nuclear matrix elements for $^{124}$Xe within the pn-QRPA with isospin restoration are presented as function of the effective value of axial-vector coupling constant $g_A^{\rm eff}$ for fixed value of  $2\nu$ECEC half-life, which corresponds to its experimental value. We see that $M^{2\nu \rm ECEC}_{GT-1}$ depend significantly on $g_A^{\rm eff}$ unlike $M^{2\nu \rm ECEC}_{GT-3}$ and $M^{2\nu \rm ECEC}_{GT-5}$. In Table \ref{tab:qrpa}, the calculated values of $M^{2\nu \rm ECEC}_{GT-K}$ (K=1, 3 and 5) obtained with the Argonne V18 and CD-Bonn potentials are displayed for unquenched and modestly quenched value of $g_A^{\rm eff}$ ($g_A^{\rm eff}$ = 1.27, 1.00 and 0.80). They are compared with corresponding nuclear matrix elements for $2\nu\beta^-\beta^-$ of $^{128}$Te and $^{130}$Te. Recall that all these transitions connect the ground states of the tellurium and xenon isotopes. By glancing at Table \ref{tab:qrpa}, we see that $ 2\nu$ECEC NMEs are only slightly larger, about up to a factor of 2, when compared with $2\nu\beta^-\beta^-$ NMEs. It reflects the fact that experimental paring gaps of all involved isotopes, which are also shown in Table \ref{tab:qrpa}, are comparable. The ratios of NMEs $\xi_{13}$ and $\xi_{15}$ are presented as well. The maximal value $\xi_{13}$ evaluated for the unquenched value of $g_A^{\rm eff}$ is 0.231, 0.282 and 0.281 in the case of $2\nu$ECEC of $^{124}$Xe, $2\nu\beta^-\beta^-$ of $^{128}$Te and $^{130}$Te and CD-Bonn potential. Recall that $\xi_{13}>0.26$ was excluded for the $2\nu\beta^-\beta^-$-decay of $^{136}$Xe by the KamLAND-Zen Collaboration \cite{KamLAND-Zen-PRL2019}. The measured value $\xi_{13} = 0.45 \pm 0.03(stat) \pm0.05(syst)$ was achieved by the CUPID-Mo experiment \cite{Augier-PRL2023}.

Before the half-life of the $2\nu$ECEC of $^{124}$Xe was measured, the nuclear matrix element for this process was calculated within the pn-QRPA by choosing a different way of adjusting the particle-particle interaction strength parameter $g_{pp}$, in particular by using available experimental information on single $\beta$-decays \cite{SuhonenJPG2013} or exploiting the formalism of statistical approach, a novel possibility to reconcile QRPA results with experimental data \cite{Faessler:2007hu,PirinenPRC2015}. Results of the second approach favor a strong quenching of axial-vector coupling constant with $g^{\rm eff}_A \simeq 0.4-0.6$. The obtained nuclear matrix elements in these studies are as follows:
$M^{2\nu \rm ECEC}_{GT} = 0.10-0.20 (g_A^{\rm eff}=1.25), 0.34-0.71 (g^{\rm eff}_A =1.00)$ \cite{SuhonenJPG2013} and
$M^{2\nu \rm ECEC}_{GT} = 0.296 (g^{\rm eff}_A =0.60)$\cite{PirinenPRC2015}. By comparing them with NMEs values in Table \ref{tab:qrpa} and with $M^{2\nu \rm ECEC}_{GT}$ = 0.191 and 0.186 (Argonne and CD-Bonn potentials) for $g^{\rm eff}_A$ =0.60 calculated in the presented formalism, we see that they are too large leading to a significant disagreement with measured half-life for $2\nu$ECEC of $^{124}$Xe.

\subsection{Total and partial half-lives ISM calculations}
 
Table \ref{hlvs} presents the total $2\nu \textrm{ECEC}$ half-life and the half-life for the KK channel calculated with Eqs. (\ref{ttt}-\ref{tkk}), obtained using the PSF of Table \ref{tab:PSFTaylor}, the ISM NMEs of Table \ref{nmes} and the ISM $\xi^{2\nu \textrm{ECEC}}_{i}$ parameters in Table \ref{xis}. Table \ref{hlvs} also includes the experimental values for comparison. The overall conclusion is that the interacting shell model results seem to be robust relative to changes of truncations and effective Hamiltonian and are in good agreement with recently reported experimental data. One should also remark that the Taylor expansion correction, as well as the newly more accurate PSF, contributed to getting the calculated half-lives closer to the experimental data.

\begin{table}[h]
\centering
\begin{tabular}{c|c|c|c|c}
Model & Channel & jump=0 & jump=2 & jump=4  \cr 
		\hline
 \hline 
SVD &  Total & 1.94 &  1.61 &  2.12     \\
&  $KK$ & 2.61 &  2.17 &  2.86     \\
\hline
GCN &  Total & 1.20 &  1.40 &  2.58    \\
&  $KK$ & 1.62 &  1.88 &  3.48    \\
\hline

\end{tabular}
\caption{\label{hlvs} The predicted $2\nu$ECEC half-lives for $^{124}$Xe (in units of $10^{22}$ yr) from Eqs. (\ref{ttt}-\ref{tkk}). To be compared with experimental data for the total half-life, $(1.1\pm0.2_{stat}\pm0.1_{sys})\times10^{22}$ yr and the inferred data for the KK half-life, $(1.5\pm0.3_{stat}\pm0.1_{sys})\times10^{22}$ yr (see section III.F of Ref. \cite{XENON1nTPRC2022}).}
\end{table}

We note, for comparison, that various estimates of the half-life of $^{124}$Xe can be found in the literature that employs various models for the NME calculation, e.g., the pn-QRPA in Refs.~\cite{SuhonenJPG2013,PirinenPRC2015} and the ISM in Ref.~\cite{COELLOPEREZPLB2019}. For example, Ref.~\cite{SuhonenJPG2013} provides several predictions for the   half-life of $^{124}$Xe for the 2$\nu$ECEC process within the range of $(0.04$--$0.88)\times 10^{22}$~yr. However, their NMEs are larger than the ones obtained here by a factor of about 5, thus making them incompatible with the most accurate PSF shown in Table I. In addition, the author refers to the expressions of~\cite{DoiPTP1992} for the calculation of PSFs, which employ the closure approximation. Similar considerations apply to Ref.~\cite{PirinenPRC2015}, which reports an estimate for the half-life within the range of $(1.4$--$1.8)\times 10^{22}$~yr. Ref.~\cite{COELLOPEREZPLB2019} reports two other ranges for the partial half-life of $^{124}$Xe, one obtained with ISM and one with an effective theory (ET) extension to truncated ISM NME. The resulting half-life expectations for the KK capture are $(1.3$--$18)\times 10^{22}$~y and $(0.43$--$2.9)\times10^{22}$~y, respectively. However, the PSF value adopted from \cite{KotilaPRC2013} also accounts for KL$_1$ and L$_1$L$_1$ captures. Consequently, the reported ranges are underestimated by about 20\%. Moreover, the PSF value employed is computed within the closure approximation while the NME is computed by summing over all intermediate $1^+$ states in SM and by assuming the SSD hypothesis in ET.

\begin{table}[h]
\centering
\begin{tabular}{c|c|c}
Decay Chanel& $R_{\rm{xy}}$ (keV) & ISM CF (\%) \cr 
		\hline
		$\rm{KK}$   & 64.62 & 74.13-74.15 \\
		$\rm{KL_{1}}$ & 37.05 & 18.76-18.83\\
		$\rm{KM_{1}}$ & 32.98 & 3.83-3.84 \\
		$\rm{KN_{1}}$ & 32.11 & 0.83-0.85 \\
		$\rm{KO_{1}}$ & 31.93 & 0.13 \\
		$\rm{L_1L_1}$ & 10.04 & 1.22 \\
		$\rm{L_1M_1}$ & 6.01 & 0.49 \\
		Other   & $ <6$  & 0.52-0.55 \\
\end{tabular}
\caption{\label{tab:RelaxationEnergies}The atomic relaxation energies (Eq.~\ref{eq:AtomicRelaxationEnergy}) obtained within the DHFS model (second column) and the capture fractions (CF) predicted by ISM (third column). The captures with atomic relaxation energies below $6$ keV are subsumed under the label "other". The ranges presented for the KK and KL$_1$ channels correspond to the minimum and maximum values of the $\xi_{31}^{2\nu\textrm{ECEC}}$ parameter predicted from ISM.}
\end{table}

Given the success in describing the experimental data for the total $2\nu$ECEC half-life of $^{124}$Xe, as well as the partial half-life for the KK channel, we make predictions for other possible channels. 
Table \ref{tab:RelaxationEnergies} presents, in addition to the relaxation energies calculated with Eq.~\ref{eq:AtomicRelaxationEnergy}, the predicted caption fractions for different channels of the decay, including the measured total and KK \cite{XENON1nTPRC2022}. The ISM CF column shows the ISM predicted capture fraction (in \%) for different channels (the detailed results for different truncations and effective Hamiltonians will be published elsewhere). 
We note that the predicted CF for the KK channel (74.1\%) differs from the value used in a previous experimental search (72.4\%)~\cite{XENON1nTPRC2022}. This difference stems from our treatment considering only $s$-wave electrons, whereas in~\cite{XENON1nTPRC2022} all electrons contribute to the process. In addition, the newly considered Taylor expansion terms and the associated PSFs have a small contribution to the new caption fractions.
In addition, we predict that the next most probable channel after KK is the $\rm{KL_{1}}$ with a CF of about 19\%. Given the closeness of the relaxation energies for the   $\rm{KL_{1}}$-$\rm{KO_{1}}$ channels, we predict a cumulative CF for those channels of about 24\%, which is about one-third of the $KK$ channel. This result suggests that the contribution from the $\rm{KL_{1}}$-$\rm{KO_{1}}$ channels could be observed in future experiments.

It is important to note that the observables associated with the $2\nu\rm{ECEC}$ process, such as the total and partial half-life, are dependent on the $\xi_{31}^{2\nu\rm{ECEC}}$ and $\xi_{51}^{2\nu\rm{ECEC}}$ parameters in the Taylor expansion formalism. From an experimental perspective, the ratio $\xi_{31}^{2\nu\rm{ECEC}}$ can be treated as a free parameter when measuring these observables, and $\xi_{51}^{2\nu\rm{ECEC}}$ can be fixed from a theoretical estimation as it has a much smaller influence. Thus, an experimental constraint of $\xi_{31}^{2\nu\rm{ECEC}}$ can be obtained from the measurement of the ratio between the half-lives of two channels, e.g. $T^{2\nu\rm{KK}}_{1/2}/T^{2\nu\rm{KL}_1}_{1/2}$, similar to the method proposed in Ref.~\cite{Nitescu-U2021} for the measurement of $\xi_{31}$ in $2\nu\beta^-\beta^-$ decays from the angular correlation coefficient of the emitted electrons. Although the half-lives ratios are rather weakly dependent on $\xi_{31}^{2\nu\textrm{ECEC}}$, a future constraint on this parameter might provide valuable insight and validation of different NMEs calculations.

 \section{Conclusions}
 \label{sec:Conclusions}

In conclusion, we studied the $2\nu $ECEC decay rates for the $^{124}$Xe isotope, that was recently investigated experimentally \cite{XENON_Collaboration2019-np,XENON1nTPRC2022}, improving the calculation of both the phase space factors and the nuclear matrix elements. The decay rate was obtained more rigorously using the Taylor expansion approach, which leads to higher-order contributions. 

In the calculation of the new PSFs, we made the following improvements: (i) we used the DHFS self-consistent framework to incorporate effects such as atomic screening, diffuse nuclear surface correction, realistic nuclear charge density, and electron exchange-correlation effects, and (ii) we considered the Pauli blocking of the decay of the innermost nucleon states. Moreover, we consider captures of all $s$-wave electrons, expanding beyond the K and L$_1$ orbitals considered in prior studies. Our findings indicate that a more precise screening correction for bound states leads to a decrease in the decay rate, while the Pauli blocking effect and the inclusion of new captures tend to increase its value.

The nuclear part was treated with ISM and pn-QRPA methods. We extended the ISM approach that calculates the sum on the $1^+$ states of the intermediate nucleus to the new Taylor expansion terms of Eqs. (\ref{ttt}-\ref{NME}). 
Additionally, we used a larger shell model truncation in the $jj55$ single particle model space, and we validated our results using two often-employed effective Hamiltonians in this single particle model space.
We calculated the new NMEs, $M_{GT-3}^{2\nu \rm{ECEC}}$ and $M_{GT-5}^{2\nu \rm{ECEC}}$, by extending the approach presented in section 4 of Ref. \cite{physics4040074}. It is worth noting that the reasonably good agreement between the ISM calculated NME and those that can be extracted from experimental data reinforces confidence in the similar prediction made for the $2\nu\beta^-\beta^-$ decay of $^{124}$Sn \cite{NeacsuHoroi2016}. 

With the ISM approach we described with good accuracy, i.e., within a factor of two, the existing experimental data for the total half-life. We also predict the capture fraction of the KK channel to be 74.1\%, which differs from the value used in past experimental searches \cite{XENON1nTPRC2022}. In addition, we made new predictions for the capture fractions for a few more specific decay channels (see Table~\ref{tab:RelaxationEnergies}) that may be accessible experimentally in the near future. Specifically, we obtained a cumulative capture fraction of about 24\%, which is about one-third of that of the KK channel. This result suggests that the contribution from the $\rm{KL_{1}}$-$\rm{KO_{1}}$ channels could be observed at relaxation energies in the interval $37.05-31.93$ keV.

The pn-QRPA with a restoration of the isospin symmetry was used to calculate nuclear matrix elements associated with the $2\nu$ECEC of $^{124}$Xe. The particle-particle strength parameter $g^{T=0}_{pp}$ was fixed by the data on the $2\nu$ECEC decay rate  of $^{124}$Xe. We found that $M_{GT-1}^{2\nu \rm{ECEC}}$ is much smaller when compared with results of previous pn-QRPA calculations \cite{SuhonenJPG2013,PirinenPRC2015}. Our results show that values of $2\nu \rm{ECEC}$ NMEs for $^{124}$Xe are comparable with those for $2\nu\beta^-\beta^-$ of $^{128,130}$Te. The corresponding nuclear systems differ only in the number of neutrons and exhibit comparable experimental pairing gaps, which constitute an important input for the pn-QRPA calculations of nuclear matrix elements.

\section*{Acknowledgements}

O.N., S.G., V.A.S., S.S., and M.H. acknowledge support from project PNRR-I8/C9-CF264, Contract No. 760100/23.05.2023 of the Romanian Ministry of Research, Innovation and Digitization. M.H. acknowledges support from the US Department of Energy grant DE-SC0022538 "Nuclear Astrophysics and Fundamental Symmetries". F.\v{S}. acknowledges support from the Slovak Research and Development Agency under Contract No. APVV-22-0413 and by the Czech Science Foundation (GA\v{C}R), project No. 24-10180S.

The authors wish to thank Dr. Andrei Neac\c{s}u for fruitful and constructive discussions during all stages of the manuscript preparation.

\biboptions{numbers,sort&compress,square}
\bibliographystyle{elsarticle-num} 
\bibliography{myPHDbibliography}

\begin{thebibliography}{10}
\expandafter\ifx\csname url\endcsname\relax
  \def\url#1{\texttt{#1}}\fi
\expandafter\ifx\csname urlprefix\endcsname\relax\def\urlprefix{URL }\fi
\expandafter\ifx\csname href\endcsname\relax
  \def\href#1#2{#2} \def\path#1{#1}\fi

\bibitem{GERDAPRL2020}
M.~Agostini, G.~R. Araujo, A.~M. Bakalyarov, M.~Balata, I.~Barabanov,
  L.~Baudis, et~al.,
  \href{https://link.aps.org/doi/10.1103/PhysRevLett.125.252502}{Final results
  of gerda on the search for neutrinoless double-$\ensuremath{\beta}$ decay},
  Phys. Rev. Lett. 125 (2020) 252502.
\newblock \href {https://doi.org/10.1103/PhysRevLett.125.252502}
  {\path{doi:10.1103/PhysRevLett.125.252502}}.
\newline\urlprefix\url{https://link.aps.org/doi/10.1103/PhysRevLett.125.252502}

\bibitem{MAJORANA2023}
I.~J. Arnquist, F.~T. Avignone, A.~S. Barabash, C.~J. Barton, P.~J. Barton,
  K.~H. Bhimani, et~al.,
  \href{https://link.aps.org/doi/10.1103/PhysRevLett.130.062501}{Final result
  of the majorana demonstrator's search for neutrinoless
  double-$\ensuremath{\beta}$ decay in $^{76}\mathrm{Ge}$}, Phys. Rev. Lett.
  130 (2023) 062501.
\newblock \href {https://doi.org/10.1103/PhysRevLett.130.062501}
  {\path{doi:10.1103/PhysRevLett.130.062501}}.
\newline\urlprefix\url{https://link.aps.org/doi/10.1103/PhysRevLett.130.062501}

\bibitem{CUPID-0PRL2022}
O.~Azzolini, J.~W. Beeman, F.~Bellini, M.~Beretta, M.~Biassoni, C.~Brofferio,
  et~al., \href{https://link.aps.org/doi/10.1103/PhysRevLett.129.111801}{Final
  result on the neutrinoless double beta decay of $^{82}\mathrm{Se}$ with
  cupid-0}, Phys. Rev. Lett. 129 (2022) 111801.
\newblock \href {https://doi.org/10.1103/PhysRevLett.129.111801}
  {\path{doi:10.1103/PhysRevLett.129.111801}}.
\newline\urlprefix\url{https://link.aps.org/doi/10.1103/PhysRevLett.129.111801}

\bibitem{CUOREPRL2022}
D.~Q. Adams, C.~Alduino, K.~Alfonso, F.~T. Avignone, O.~Azzolini, G.~Bari,
  et~al., \href{https://link.aps.org/doi/10.1103/PhysRevLett.129.222501}{New
  direct limit on neutrinoless double beta decay half-life of
  $^{128}\mathrm{Te}$ with cuore}, Phys. Rev. Lett. 129 (2022) 222501.
\newblock \href {https://doi.org/10.1103/PhysRevLett.129.222501}
  {\path{doi:10.1103/PhysRevLett.129.222501}}.
\newline\urlprefix\url{https://link.aps.org/doi/10.1103/PhysRevLett.129.222501}

\bibitem{SNO2022}
J.~R. Klein, T.~Akindele, A.~Bernstein, S.~Biller, N.~Bowden, J.~Brodsky,
  et~al., Future advances in photon-based neutrino detectors: A snowmass white
  paper (2022).
\newblock \href {http://arxiv.org/abs/2203.07479} {\path{arXiv:2203.07479}}.

\bibitem{KamLAND-ZenPRL2023}
S.~Abe, S.~Asami, M.~Eizuka, S.~Futagi, A.~Gando, Y.~Gando, et~al.,
  \href{https://link.aps.org/doi/10.1103/PhysRevLett.130.051801}{Search for the
  majorana nature of neutrinos in the inverted mass ordering region with
  kamland-zen}, Phys. Rev. Lett. 130 (2023) 051801.
\newblock \href {https://doi.org/10.1103/PhysRevLett.130.051801}
  {\path{doi:10.1103/PhysRevLett.130.051801}}.
\newline\urlprefix\url{https://link.aps.org/doi/10.1103/PhysRevLett.130.051801}

\bibitem{LEGEND2021}
N.~Abgrall, I.~Abt, M.~Agostini, A.~Alexander, C.~Andreoiu, G.~R. Araujo,
  et~al., Legend-1000 preconceptual design report (2021).
\newblock \href {http://arxiv.org/abs/2107.11462} {\path{arXiv:2107.11462}}.

\bibitem{VergadosRPR2012}
J.~D. Vergados, H.~Ejiri, F.~Šimkovic,
  \href{https://dx.doi.org/10.1088/0034-4885/75/10/106301}{Theory of
  neutrinoless double-beta decay}, Reports on Progress in Physics 75~(10)
  (2012) 106301.
\newblock \href {https://doi.org/10.1088/0034-4885/75/10/106301}
  {\path{doi:10.1088/0034-4885/75/10/106301}}.
\newline\urlprefix\url{https://dx.doi.org/10.1088/0034-4885/75/10/106301}

\bibitem{DolinskiARNPS2019}
M.~J. Dolinski, A.~W. Poon, W.~Rodejohann,
  \href{https://doi.org/10.1146/annurev-nucl-101918-023407}{Neutrinoless
  double-beta decay: Status and prospects}, Annual Review of Nuclear and
  Particle Science 69~(1) (2019) 219--251.
\newblock \href
  {http://arxiv.org/abs/https://doi.org/10.1146/annurev-nucl-101918-023407}
  {\path{arXiv:https://doi.org/10.1146/annurev-nucl-101918-023407}}, \href
  {https://doi.org/10.1146/annurev-nucl-101918-023407}
  {\path{doi:10.1146/annurev-nucl-101918-023407}}.
\newline\urlprefix\url{https://doi.org/10.1146/annurev-nucl-101918-023407}

\bibitem{AdamsARXIV2022}
C.~Adams, K.~Alfonso, C.~Andreoiu, E.~Angelico, I.~J. Arnquist, J.~A.~A.
  Asaadi, et~al., Neutrinoless double beta decay (2022).
\newblock \href {http://arxiv.org/abs/2212.11099} {\path{arXiv:2212.11099}}.

\bibitem{BlaumRMP2020}
K.~Blaum, S.~Eliseev, F.~A. Danevich, V.~I. Tretyak, S.~Kovalenko, M.~I.
  Krivoruchenko, Y.~N. Novikov, J.~Suhonen,
  \href{https://link.aps.org/doi/10.1103/RevModPhys.92.045007}{Neutrinoless
  double-electron capture}, Rev. Mod. Phys. 92 (2020) 045007.
\newblock \href {https://doi.org/10.1103/RevModPhys.92.045007}
  {\path{doi:10.1103/RevModPhys.92.045007}}.
\newline\urlprefix\url{https://link.aps.org/doi/10.1103/RevModPhys.92.045007}

\bibitem{Belli-P2021}
P.~Belli, R.~Bernabei, V.~Caracciolo,
  \href{https://www.mdpi.com/2571-712X/4/2/23}{Status and perspectives of
  2$\epsilon$, $\epsilon\beta+$ and 2$\beta$ decays}, Particles 4~(2) (2021)
  241--274.
\newblock \href {https://doi.org/10.3390/particles4020023}
  {\path{doi:10.3390/particles4020023}}.
\newline\urlprefix\url{https://www.mdpi.com/2571-712X/4/2/23}

\bibitem{xmass-2018}
X.~Collaboration, K.~Abe, K.~Hiraide, K.~Ichimura, Y.~Kishimoto, K.~Kobayashi,
  M.~Kobayashi, S.~Moriyama, M.~Nakahata, T.~Norita, H.~Ogawa, K.~Sato,
  H.~Sekiya, O.~Takachio, A.~Takeda, S.~Tasaka, M.~Yamashita, B.~S. Yang, N.~Y.
  Kim, Y.~D. Kim, Y.~Itow, K.~Kanzawa, R.~Kegasa, K.~Masuda, H.~Takiya,
  K.~Fushimi, G.~Kanzaki, K.~Martens, Y.~Suzuki, B.~D. Xu, R.~Fujita,
  K.~Hosokawa, K.~Miuchi, N.~Oka, Y.~Takeuchi, Y.~H. Kim, K.~B. Lee, M.~K. Lee,
  Y.~Fukuda, M.~Miyasaka, K.~Nishijima, S.~Nakamura,
  \href{https://doi.org/10.1093/ptep/pty053}{{Improved search for two-neutrino
  double electron capture on 124Xe and 126Xe using particle identification in
  XMASS-I}}, Progress of Theoretical and Experimental Physics 2018~(5) (2018)
  053D03.
\newblock \href
  {http://arxiv.org/abs/https://academic.oup.com/ptep/article-pdf/2018/5/053D03/24969810/pty053.pdf}
  {\path{arXiv:https://academic.oup.com/ptep/article-pdf/2018/5/053D03/24969810/pty053.pdf}},
  \href {https://doi.org/10.1093/ptep/pty053} {\path{doi:10.1093/ptep/pty053}}.
\newline\urlprefix\url{https://doi.org/10.1093/ptep/pty053}

\bibitem{XENONnT2019}
E.~Aprile, J.~Aalbers, F.~Agostini, M.~Alfonsi, L.~Althueser, F.~D. Amaro,
  et~al., \href{https://doi.org/10.1038/s41586-019-1124-4}{Observation of
  two-neutrino double electron capture in 124xe with xenon1t}, Nature
  568~(7753) (2019) 532--535.
\newblock \href {https://doi.org/10.1038/s41586-019-1124-4}
  {\path{doi:10.1038/s41586-019-1124-4}}.
\newline\urlprefix\url{https://doi.org/10.1038/s41586-019-1124-4}

\bibitem{XENON1nTPRC2022}
E.~Aprile, K.~Abe, F.~Agostini, S.~Ahmed~Maouloud, M.~Alfonsi, L.~Althueser,
  et~al.,
  \href{https://link.aps.org/doi/10.1103/PhysRevC.106.024328}{Double-weak
  decays of $^{124}\mathrm{Xe}$ and $^{136}\mathrm{Xe}$ in the xenon1t and
  xenonnt experiments}, Phys. Rev. C 106 (2022) 024328.
\newblock \href {https://doi.org/10.1103/PhysRevC.106.024328}
  {\path{doi:10.1103/PhysRevC.106.024328}}.
\newline\urlprefix\url{https://link.aps.org/doi/10.1103/PhysRevC.106.024328}

\bibitem{NEXT2021}
G.~Mart{\'i}nez-Lema, M.~Mart{\'i}nez-Vara, M.~Sorel, C.~Adams, V.~{\'A}lvarez,
  L.~Arazi, I.~J. Arnquist, et~al.,
  \href{https://doi.org/10.1007/JHEP02(2021)203}{Sensitivity of the next
  experiment to xe-124 double electron capture}, Journal of High Energy Physics
  2021~(2) (2021) 203.
\newblock \href {https://doi.org/10.1007/JHEP02(2021)203}
  {\path{doi:10.1007/JHEP02(2021)203}}.
\newline\urlprefix\url{https://doi.org/10.1007/JHEP02(2021)203}

\bibitem{LUX2020}
D.~S. Akerib, S.~Alsum, H.~M. Araújo, X.~Bai, J.~Balajthy, A.~Baxter, et~al.,
  \href{https://dx.doi.org/10.1088/1361-6471/ab9c2d}{Search for two neutrino
  double electron capture of 124xe and 126xe in the full exposure of the lux
  detector}, Journal of Physics G: Nuclear and Particle Physics 47~(10) (2020)
  105105.
\newblock \href {https://doi.org/10.1088/1361-6471/ab9c2d}
  {\path{doi:10.1088/1361-6471/ab9c2d}}.
\newline\urlprefix\url{https://dx.doi.org/10.1088/1361-6471/ab9c2d}

\bibitem{SuhonenJPG2013}
J.~Suhonen, \href{https://dx.doi.org/10.1088/0954-3899/40/7/075102}{Double beta
  decays of 124xe investigated in the qrpa framework}, Journal of Physics G:
  Nuclear and Particle Physics 40~(7) (2013) 075102.
\newblock \href {https://doi.org/10.1088/0954-3899/40/7/075102}
  {\path{doi:10.1088/0954-3899/40/7/075102}}.
\newline\urlprefix\url{https://dx.doi.org/10.1088/0954-3899/40/7/075102}

\bibitem{PirinenPRC2015}
P.~Pirinen, J.~Suhonen,
  \href{https://link.aps.org/doi/10.1103/PhysRevC.91.054309}{Systematic
  approach to $\ensuremath{\beta}$ and
  $2\ensuremath{\nu}\ensuremath{\beta}\ensuremath{\beta}$ decays of mass
  $a=100--136$ nuclei}, Phys. Rev. C 91 (2015) 054309.
\newblock \href {https://doi.org/10.1103/PhysRevC.91.054309}
  {\path{doi:10.1103/PhysRevC.91.054309}}.
\newline\urlprefix\url{https://link.aps.org/doi/10.1103/PhysRevC.91.054309}

\bibitem{COELLOPEREZPLB2019}
E.~{Coello Pérez}, J.~Menéndez, A.~Schwenk,
  \href{https://www.sciencedirect.com/science/article/pii/S0370269319305994}{Two-neutrino
  double electron capture on 124xe based on an effective theory and the nuclear
  shell model}, Physics Letters B 797 (2019) 134885.
\newblock \href
  {https://doi.org/https://doi.org/10.1016/j.physletb.2019.134885}
  {\path{doi:https://doi.org/10.1016/j.physletb.2019.134885}}.
\newline\urlprefix\url{https://www.sciencedirect.com/science/article/pii/S0370269319305994}

\bibitem{DoiPTP1992}
M.~Doi, T.~Kotani, \href{https://doi.org/10.1143/ptp/87.5.1207}{{Neutrino
  Emitting Modes of Double Beta Decay}}, Progress of Theoretical Physics 87~(5)
  (1992) 1207--1231.
\newblock \href {https://doi.org/10.1143/ptp/87.5.1207}
  {\path{doi:10.1143/ptp/87.5.1207}}.
\newline\urlprefix\url{https://doi.org/10.1143/ptp/87.5.1207}

\bibitem{Simkovic-2018}
F.~\ifmmode~\check{S}\else \v{S}\fi{}imkovic, R.~Dvornick\'y, D.~c.~v.
  \ifmmode~\check{S}\else \v{S}\fi{}tef\'anik, A.~Faessler, Improved
  description of the
  $2\ensuremath{\nu}\ensuremath{\beta}\ensuremath{\beta}$-decay and a
  possibility to determine the effective axial-vector coupling constant, Phys.
  Rev. C 97 (2018) 034315.

\bibitem{Nitescu-U2021}
O.~Ni\c{t}escu, R.~Dvornick\'y, S.~Stoica, F.~\v{S}imkovic,
  \href{https://www.mdpi.com/2218-1997/7/5/147}{Angular distributions of
  emitted electrons in the two-neutrino $\beta\beta$ decay}, Universe 7~(5)
  (2021).
\newblock \href {https://doi.org/10.3390/universe7050147}
  {\path{doi:10.3390/universe7050147}}.
\newline\urlprefix\url{https://www.mdpi.com/2218-1997/7/5/147}

\bibitem{KotilaPRC2013}
J.~Kotila, F.~Iachello,
  \href{https://link.aps.org/doi/10.1103/PhysRevC.87.024313}{Phase space
  factors for ${\ensuremath{\beta}}^{+}{\ensuremath{\beta}}^{+}$ decay and
  competing modes of double-$\ensuremath{\beta}$ decay}, Phys. Rev. C 87 (2013)
  024313.
\newblock \href {https://doi.org/10.1103/PhysRevC.87.024313}
  {\path{doi:10.1103/PhysRevC.87.024313}}.
\newline\urlprefix\url{https://link.aps.org/doi/10.1103/PhysRevC.87.024313}

\bibitem{StoicaFP2019}
S.~Stoica, M.~Mirea,
  \href{https://www.frontiersin.org/articles/10.3389/fphy.2019.00012}{Phase
  space factors for double-beta decays}, Frontiers in Physics 7 (2019).
\newblock \href {https://doi.org/10.3389/fphy.2019.00012}
  {\path{doi:10.3389/fphy.2019.00012}}.
\newline\urlprefix\url{https://www.frontiersin.org/articles/10.3389/fphy.2019.00012}

\bibitem{HoroiStoicaBrown2007}
M.~Horoi, S.~Stoica, B.~A. Brown, Shell-model calculations of two-neutrino
  double-beta decay rates of $^{48}${Ca} with the {GXPF1A} interaction, Phys.
  Rev. C 75 (2007) 034303.
\newblock \href {https://doi.org/10.1103/PhysRevC.75.034303}
  {\path{doi:10.1103/PhysRevC.75.034303}}.

\bibitem{SenkovHoroiBrown2014}
R.~A. Sen'kov, M.~Horoi, B.~A. Brown, Neutrinoless double-beta decay of se-82
  in the shell model: Beyond the closure approximation, Phys. Rev. C 89~(5)
  (2014) 054304.
\newblock \href {https://doi.org/10.1103/PhysRevC.89.054304}
  {\path{doi:10.1103/PhysRevC.89.054304}}.

\bibitem{SenkovHoroi2016}
R.~A. Sen'kov, M.~Horoi, Shell-model calculation of neutrinoless
  double-$\ensuremath{\beta}$ decay of $^{76}\mathrm{Ge}$, Phys. Rev. C 93
  (2016) 044334.
\newblock \href {https://doi.org/10.1103/PhysRevC.93.044334}
  {\path{doi:10.1103/PhysRevC.93.044334}}.

\bibitem{NeacsuHoroi2015}
A.~Neacsu, M.~Horoi, Shell model studies of the $^{130}${Te} neutrinoless
  double-beta decay, Phys. Rev. C 91 (2015) 024309.
\newblock \href {https://doi.org/10.1103/PhysRevC.91.024309}
  {\path{doi:10.1103/PhysRevC.91.024309}}.

\bibitem{NeacsuHoroi2016}
M.~Horoi, A.~Neacsu, {Shell model predictions for $^{124}$Sn double-$\beta$
  decay}, Phys. Rev. C 93 (2016) 024308.
\newblock \href {https://doi.org/10.1103/PhysRevC.93.024308}
  {\path{doi:10.1103/PhysRevC.93.024308}}.

\bibitem{physics4040074}
M.~Horoi, \href{https://www.mdpi.com/2624-8174/4/4/74}{Double beta decay: A
  shell model approach}, Physics 4~(4) (2022) 1135--1149.
\newblock \href {https://doi.org/10.3390/physics4040074}
  {\path{doi:10.3390/physics4040074}}.
\newline\urlprefix\url{https://www.mdpi.com/2624-8174/4/4/74}

\bibitem{Sim13}
F.~\ifmmode~\check{S}\else \v{S}\fi{}imkovic, V.~Rodin, A.~Faessler, P.~Vogel,
  \href{https://link.aps.org/doi/10.1103/PhysRevC.87.045501}{$0\ensuremath{\nu}\ensuremath{\beta}\ensuremath{\beta}$
  and $2\ensuremath{\nu}\ensuremath{\beta}\ensuremath{\beta}$ nuclear matrix
  elements, quasiparticle random-phase approximation, and isospin symmetry
  restoration}, Phys. Rev. C 87 (2013) 045501.
\newblock \href {https://doi.org/10.1103/PhysRevC.87.045501}
  {\path{doi:10.1103/PhysRevC.87.045501}}.
\newline\urlprefix\url{https://link.aps.org/doi/10.1103/PhysRevC.87.045501}

\bibitem{Angeli-ADNDT2013}
I.~Angeli, K.~Marinova,
  \href{https://www.sciencedirect.com/science/article/pii/S0092640X12000265}{Table
  of experimental nuclear ground state charge radii: An update}, Atomic Data
  and Nuclear Data Tables 99~(1) (2013) 69--95.
\newblock \href {https://doi.org/https://doi.org/10.1016/j.adt.2011.12.006}
  {\path{doi:https://doi.org/10.1016/j.adt.2011.12.006}}.
\newline\urlprefix\url{https://www.sciencedirect.com/science/article/pii/S0092640X12000265}

\bibitem{NitescuPRC2023}
O.~Ni\ifmmode~\mbox{\c{t}}\else \c{t}\fi{}escu, S.~Stoica,
  F.~\ifmmode~\check{S}\else \v{S}\fi{}imkovic,
  \href{https://link.aps.org/doi/10.1103/PhysRevC.107.025501}{Exchange
  correction for allowed $\ensuremath{\beta}$ decay}, Phys. Rev. C 107 (2023)
  025501.
\newblock \href {https://doi.org/10.1103/PhysRevC.107.025501}
  {\path{doi:10.1103/PhysRevC.107.025501}}.
\newline\urlprefix\url{https://link.aps.org/doi/10.1103/PhysRevC.107.025501}

\bibitem{SevestreanPRA2023}
V.~A. Sevestrean, O.~Ni\ifmmode~\mbox{\c{t}}\else \c{t}\fi{}escu, S.~Ghinescu,
  S.~Stoica,
  \href{https://link.aps.org/doi/10.1103/PhysRevA.108.012810}{Self-consistent
  calculations for atomic electron capture}, Phys. Rev. A 108 (2023) 012810.
\newblock \href {https://doi.org/10.1103/PhysRevA.108.012810}
  {\path{doi:10.1103/PhysRevA.108.012810}}.
\newline\urlprefix\url{https://link.aps.org/doi/10.1103/PhysRevA.108.012810}

\bibitem{Salvat-CPC2019}
F.~Salvat, J.~M. Fernández-Varea,
  \href{https://www.sciencedirect.com/science/article/pii/S0010465519300633}{{RADIAL}:
  A fortran subroutine package for the solution of the radial {S}chr\"odinger
  and {D}irac wave equations}, Computer Physics Communications 240 (2019)
  165--177.
\newblock \href {https://doi.org/https://doi.org/10.1016/j.cpc.2019.02.011}
  {\path{doi:https://doi.org/10.1016/j.cpc.2019.02.011}}.
\newline\urlprefix\url{https://www.sciencedirect.com/science/article/pii/S0010465519300633}

\bibitem{Nitescu-U2024}
O.~Niţescu, S.~Ghinescu, S.~Stoica, F.~Šimkovic,
  \href{https://www.mdpi.com/2218-1997/10/2/98}{A systematic study of
  two-neutrino double electron capture}, Universe 10~(2) (2024).
\newblock \href {https://doi.org/10.3390/universe10020098}
  {\path{doi:10.3390/universe10020098}}.
\newline\urlprefix\url{https://www.mdpi.com/2218-1997/10/2/98}

\bibitem{QValue124Xe-PRC2012}
D.~A. Nesterenko, K.~Blaum, M.~Block, C.~Droese, S.~Eliseev, F.~Herfurth,
  E.~Minaya~Ramirez, Y.~N. Novikov, L.~Schweikhard, V.~M. Shabaev, M.~V.
  Smirnov, I.~I. Tupitsyn, K.~Zuber, N.~A. Zubova,
  \href{https://link.aps.org/doi/10.1103/PhysRevC.86.044313}{Double-$\ensuremath{\beta}$
  transformations in isobaric triplets with mass numbers
  $\mathbf{A}=\mathbf{124}$, 130, and 136}, Phys. Rev. C 86 (2012) 044313.
\newblock \href {https://doi.org/10.1103/PhysRevC.86.044313}
  {\path{doi:10.1103/PhysRevC.86.044313}}.
\newline\urlprefix\url{https://link.aps.org/doi/10.1103/PhysRevC.86.044313}

\bibitem{DARWIN-JCAP2016}
J.~Aalbers, F.~Agostini, M.~Alfonsi, F.~Amaro, C.~Amsler, E.~Aprile, et~al.,
  \href{https://dx.doi.org/10.1088/1475-7516/2016/11/017}{Darwin: towards the
  ultimate dark matter detector}, Journal of Cosmology and Astroparticle
  Physics 2016~(11) (2016) 017.
\newblock \href {https://doi.org/10.1088/1475-7516/2016/11/017}
  {\path{doi:10.1088/1475-7516/2016/11/017}}.
\newline\urlprefix\url{https://dx.doi.org/10.1088/1475-7516/2016/11/017}

\bibitem{DARWIN-JPG2022}
J.~Aalbers, S.~S. AbdusSalam, K.~Abe, V.~Aerne, F.~Agostini, S.~A. Maouloud,
  et~al., \href{https://dx.doi.org/10.1088/1361-6471/ac841a}{A next-generation
  liquid xenon observatory for dark matter and neutrino physics}, Journal of
  Physics G: Nuclear and Particle Physics 50~(1) (2022) 013001.
\newblock \href {https://doi.org/10.1088/1361-6471/ac841a}
  {\path{doi:10.1088/1361-6471/ac841a}}.
\newline\urlprefix\url{https://dx.doi.org/10.1088/1361-6471/ac841a}

\bibitem{Chong2012}
C.~Qi, Z.~X. Xu, Monopole-optimized effective interaction for tin isotopes,
  Phys. Rev. C 86~(4) (2012) 044323.
\newblock \href {https://doi.org/10.1103/PhysRevC.86.044323}
  {\path{doi:10.1103/PhysRevC.86.044323}}.

\bibitem{PhysRevC.100.014316}
L.~Coraggio, L.~De~Angelis, T.~Fukui, A.~Gargano, N.~Itaco, F.~Nowacki,
  \href{https://link.aps.org/doi/10.1103/PhysRevC.100.014316}{Renormalization
  of the gamow-teller operator within the realistic shell model}, Phys. Rev. C
  100 (2019) 014316.
\newblock \href {https://doi.org/10.1103/PhysRevC.100.014316}
  {\path{doi:10.1103/PhysRevC.100.014316}}.
\newline\urlprefix\url{https://link.aps.org/doi/10.1103/PhysRevC.100.014316}

\bibitem{particles4040038}
S.~R. Stroberg, \href{https://www.mdpi.com/2571-712X/4/4/38}{Beta decay in
  medium-mass nuclei with the in-medium similarity renormalization group},
  Particles 4~(4) (2021) 521--535.
\newblock \href {https://doi.org/10.3390/particles4040038}
  {\path{doi:10.3390/particles4040038}}.
\newline\urlprefix\url{https://www.mdpi.com/2571-712X/4/4/38}

\bibitem{Gysbers2019-tm}
P.~Gysbers, G.~Hagen, J.~D. Holt, G.~R. Jansen, T.~D. Morris, P.~Navr{\'a}til,
  T.~Papenbrock, S.~Quaglioni, A.~Schwenk, S.~R. Stroberg, K.~A. Wendt,
  Discrepancy between experimental and theoretical $\beta$-decay rates resolved
  from first principles, Nature Physics 15~(5) (2019) 428--431.

\bibitem{PhysRevLett.122.242501}
B.~M\"arkisch, H.~Mest, H.~Saul, X.~Wang, H.~Abele, D.~Dubbers, M.~Klopf,
  A.~Petoukhov, C.~Roick, T.~Soldner, D.~Werder,
  \href{https://link.aps.org/doi/10.1103/PhysRevLett.122.242501}{Measurement of
  the weak axial-vector coupling constant in the decay of free neutrons using a
  pulsed cold neutron beam}, Phys. Rev. Lett. 122 (2019) 242501.
\newblock \href {https://doi.org/10.1103/PhysRevLett.122.242501}
  {\path{doi:10.1103/PhysRevLett.122.242501}}.
\newline\urlprefix\url{https://link.aps.org/doi/10.1103/PhysRevLett.122.242501}

\bibitem{Sim04}
F.~\v{S}imkovic, L.~Pacearescu, A.~Faessler,
  \href{https://www.sciencedirect.com/science/article/pii/S037594740400003X}{Two-neutrino
  double beta decay of 76ge within deformed qrpa}, Nuclear Physics A 733~(3)
  (2004) 321--350.
\newblock \href
  {https://doi.org/https://doi.org/10.1016/j.nuclphysa.2004.01.002}
  {\path{doi:https://doi.org/10.1016/j.nuclphysa.2004.01.002}}.
\newline\urlprefix\url{https://www.sciencedirect.com/science/article/pii/S037594740400003X}

\bibitem{Barabash19}
A.~Barabash, \href{https://www.mdpi.com/2218-1997/6/10/159}{Precise half-life
  values for two-neutrino double-$\beta$ decay: 2020 review}, Universe 6~(10)
  (2020).
\newblock \href {https://doi.org/10.3390/universe6100159}
  {\path{doi:10.3390/universe6100159}}.
\newline\urlprefix\url{https://www.mdpi.com/2218-1997/6/10/159}

\bibitem{Cuore23}
D.~Q. Adams, C.~Alduino, K.~Alfonso, F.~T. Avignone, O.~Azzolini, G.~Bari,
  F.~Bellini, G.~Benato, M.~Biassoni, A.~Branca, C.~Brofferio, C.~Bucci,
  J.~Camilleri, A.~Caminata, A.~Campani, L.~Canonica, X.~G. Cao, S.~Capelli,
  L.~Cappelli, L.~Cardani, P.~Carniti, N.~Casali, D.~Chiesa, M.~Clemenza,
  S.~Copello, C.~Cosmelli, O.~Cremonesi, R.~J. Creswick, A.~D'Addabbo,
  I.~Dafinei, C.~J. Davis, S.~Dell'Oro, S.~Di~Domizio, V.~Domp\`e, D.~Q. Fang,
  G.~Fantini, M.~Faverzani, E.~Ferri, F.~Ferroni, E.~Fiorini, M.~A. Franceschi,
  S.~J. Freedman, S.~H. Fu, B.~K. Fujikawa, A.~Giachero, L.~Gironi,
  A.~Giuliani, P.~Gorla, C.~Gotti, T.~D. Gutierrez, K.~Han, K.~M. Heeger, R.~G.
  Huang, H.~Z. Huang, J.~Johnston, G.~Keppel, Y.~G. Kolomensky, C.~Ligi, L.~Ma,
  Y.~G. Ma, L.~Marini, R.~H. Maruyama, D.~Mayer, Y.~Mei, N.~Moggi, S.~Morganti,
  T.~Napolitano, M.~Nastasi, J.~Nikkel, C.~Nones, E.~B. Norman, A.~Nucciotti,
  I.~Nutini, T.~O'Donnell, J.~L. Ouellet, S.~Pagan, C.~E. Pagliarone,
  L.~Pagnanini, M.~Pallavicini, L.~Pattavina, M.~Pavan, G.~Pessina,
  V.~Pettinacci, C.~Pira, S.~Pirro, S.~Pozzi, E.~Previtali, A.~Puiu,
  C.~Rosenfeld, C.~Rusconi, M.~Sakai, S.~Sangiorgio, B.~Schmidt, N.~D. Scielzo,
  V.~Sharma, V.~Singh, M.~Sisti, D.~Speller, P.~T. Surukuchi, L.~Taffarello,
  F.~Terranova, C.~Tomei, K.~J. Vetter, M.~Vignati, S.~L. Wagaarachchi, B.~S.
  Wang, B.~Welliver, J.~Wilson, K.~Wilson, L.~A. Winslow, S.~Zimmermann,
  S.~Zucchelli,
  \href{https://link.aps.org/doi/10.1103/PhysRevLett.126.171801}{Measurement of
  the $2\ensuremath{\nu}\ensuremath{\beta}\ensuremath{\beta}$ decay half-life
  of $^{130}\mathrm{Te}$ with cuore}, Phys. Rev. Lett. 126 (2021) 171801.
\newblock \href {https://doi.org/10.1103/PhysRevLett.126.171801}
  {\path{doi:10.1103/PhysRevLett.126.171801}}.
\newline\urlprefix\url{https://link.aps.org/doi/10.1103/PhysRevLett.126.171801}

\bibitem{KamLAND-Zen-PRL2019}
A.~Gando, Y.~Gando, T.~Hachiya, M.~Ha~Minh, S.~Hayashida, Y.~Honda, et~al.,
  \href{https://link.aps.org/doi/10.1103/PhysRevLett.122.192501}{Precision
  analysis of the $^{136}\mathrm{Xe}$ two-neutrino
  $\ensuremath{\beta}\ensuremath{\beta}$ spectrum in kamland-zen and its impact
  on the quenching of nuclear matrix elements}, Phys. Rev. Lett. 122 (2019)
  192501.
\newblock \href {https://doi.org/10.1103/PhysRevLett.122.192501}
  {\path{doi:10.1103/PhysRevLett.122.192501}}.
\newline\urlprefix\url{https://link.aps.org/doi/10.1103/PhysRevLett.122.192501}

\bibitem{Augier-PRL2023}
C.~Augier, A.~S. Barabash, F.~Bellini, G.~Benato, M.~Beretta, L.~Berg\'e,
  et~al.,
  \href{https://link.aps.org/doi/10.1103/PhysRevLett.131.162501}{Measurement of
  the $2\ensuremath{\nu}\ensuremath{\beta}\ensuremath{\beta}$ decay rate and
  spectral shape of $^{100}\mathrm{Mo}$ from the cupid-mo experiment}, Phys.
  Rev. Lett. 131 (2023) 162501.
\newblock \href {https://doi.org/10.1103/PhysRevLett.131.162501}
  {\path{doi:10.1103/PhysRevLett.131.162501}}.
\newline\urlprefix\url{https://link.aps.org/doi/10.1103/PhysRevLett.131.162501}

\bibitem{Faessler:2007hu}
A.~Faessler, G.~L. Fogli, E.~Lisi, V.~Rodin, A.~M. Rotunno, F.~Šimkovic,
  \href{https://dx.doi.org/10.1088/0954-3899/35/7/075104}{Overconstrained
  estimates of neutrinoless double beta decay within the qrpa}, Journal of
  Physics G: Nuclear and Particle Physics 35~(7) (2008) 075104.
\newblock \href {https://doi.org/10.1088/0954-3899/35/7/075104}
  {\path{doi:10.1088/0954-3899/35/7/075104}}.
\newline\urlprefix\url{https://dx.doi.org/10.1088/0954-3899/35/7/075104}

\bibitem{XENON_Collaboration2019-np}
E.~Aprile, J.~Aalbers, F.~Agostini, M.~Alfonsi, L.~Althueser, F.~D. Amaro,
  M.~Anthony, V.~C. Antochi, F.~Arneodo, L.~Baudis, B.~Bauermeister, et~al.,
  Observation of two-neutrino double electron capture in ${}^{124}$xe with
  xenon1t, Nature 568~(7753) (2019) 532--535.
\newblock \href {https://doi.org/10.1038/s41586-019-1124-4}
  {\path{doi:10.1038/s41586-019-1124-4}}.

\end{thebibliography}






\end{document}